\documentclass[a4paper, 12pt]{article}
\usepackage[top=5cm, bottom=5cm, left=3cm, right=3cm]{geometry}
\usepackage{geometry}
\usepackage{amsmath, amsthm, amsfonts, multirow, float, moreverb}
\usepackage{amssymb}
\usepackage{graphicx, verbatim}
\usepackage{epsfig}
\usepackage{epstopdf}
\usepackage{subcaption}
\usepackage{parskip}
\usepackage{pdflscape}
\usepackage{pdfpages}
\usepackage{tabularx}
\usepackage{tikz}
\usepackage{longtable}
\usepackage[flushleft]{threeparttable}
\usepackage{booktabs}
\usepackage{authblk}
\usepackage{caption}
\usepackage[singlelinecheck=false]{caption}
\usepackage[font=small,labelfont=bf]{caption}
\usepackage{subcaption}
\usepackage{changes}
\usepackage{enumerate}
\usepackage[noadjust]{cite}
\usepackage{hhline}
\usepackage{makecell}
\usepackage{setspace}
\def\ci{\perp\!\!\!\perp}
\usepackage{moresize}
\providecommand{\keywords}[1]{\textbf{\text{Keywords:}} #1}
\usepackage[toc,page]{appendix}
\usepackage{abstract}

\begin{document}

\title{Estimating cluster-level local average treatment effects in cluster randomised trials with non-adherence}
\author[1]{Schadrac C. Agbla}
\author[2]{Bianca De Stavola}
\author[1]{Karla DiazOrdaz\footnote{corresponding author}}
\affil[1]{\footnotesize{Department of Medical Statistics, London School of Hygiene and Tropical Medicine, UK}}
\affil[2]{\footnotesize{Faculty of Population Health Sciences, UCL GOS Institute of Child Health, UK}}
\date{}
\maketitle
	
\begin{abstract}
Non-adherence to assigned treatment is a common issue in cluster randomised trials (CRTs). In these settings, the efficacy estimand may be also of interest. Many methodological contributions in recent years have advocated using instrumental variables to identify and estimate the local average treatment effect (LATE).
However, the clustered nature of randomisation in CRTs adds to the complexity of such analyses.

In this paper, we show that under certain assumptions, the LATE can be estimated via two-stage least squares (TSLS) using cluster-level summaries of outcomes and treatment received. Implementation needs to account for this, as well as the possible heteroscedasticity, to obtain valid inferences.

We use simulations to assess the performance of TSLS of cluster-level summaries under cluster-level or individual-level non-adherence, with and without weighting and robust standard errors. We also explore the impact of adjusting for cluster-level covariates and of appropriate degrees of freedom correction for inference.

We find that TSLS estimation using cluster-level summaries provides estimates with small to negligible bias and coverage close to nominal level, provided small sample degrees of freedom correction is used for inference, with appropriate use of robust standard errors.
We illustrate the methods by re-analysing a CRT in UK primary health settings.
\end{abstract}

\section{Introduction}
Cluster randomised trials (CRTs), which randomise groups of individuals, are common in public health and primary care. The adoption of this design is  often justified given the reduction of ``cross-over contamination'' between the experimental arms and improved adherence with allocated treatment \cite{donner2004pitfalls,glynn2007design,wright2015review}. Nevertheless, treatment non-adherence  is as common in CRTs as it is in individually randomised trials \cite{agbla2017adherence}. Dealing with non-adherence is  more challenging because there are at least two levels at which deviations from protocol can occur, e.g. cluster or individual level \cite{schochet2011estimation}.  We say that adherence is at the cluster-level if all individuals within a cluster receive the treatment the cluster was randomised to. In contrast, we say that adherence is at the individual-level, if  the treatment received varies across individuals within the same cluster,  so that some individuals received the treatment allocated to their cluster,  while others did not.

The standard analysis of randomised clinical trials is intention-to-treat (ITT), which compares  average outcomes  across randomised groups. However, if the  effect of treatment received  is confounded, in the sense that there are measured and unmeasured common causes of   receiving treatment  and  experiencing the outcome,  the ITT provides the causal effect of being offered, rather than of receiving, the treatment.
An ITT analysis with poor adherence may dilute a true treatment effect  \cite{hernan2017per}.
Recently, there has been an increased interest in estimating  other  estimands alongside the ITT, as highlighted by the International Council for Harmonisation (ICH) addendum to guideline E9 (Statistical Principles for Clinical Trials).  Amongst them, the causal effect in those adhering to treatment has been singled out as being of interest for patients  \cite{Akacha2017}.

In the presence of unmeasured confounding, instrumental variable (IV) methods  can estimate consistently the causal effect of an exposure under certain assumptions \cite{imbens1994identification,  angrist1996identification}.
An IV is a variable which is correlated with the exposure but  is not associated with any  confounders of the exposure--outcome association, nor is there any pathway by which the IV affects the outcome, other than through the exposure.
 
Since randomised treatment is usually a valid instrument, IV methods  have been proposed to estimate the treatment causal effect in the context of  randomised clinical trials  affected by  non-adherence \cite{White2005, bellamy2007introduction}.
 The population to which an IV estimate applies however, depends on the assumed behaviour of the instrument \cite{angrist1996identification}. When, as it is often the case, randomised treatment influences treatment received  \emph{monotonically} \cite{angrist1996identification}, in the sense that the level of treatment received is greater when randomised to treatment, than when randomised to the control (the precise technical definition will be given shortly), IV methods lead
to estimating the causal effect  among the adherers,  known as the local average treatment effect (LATE)  or  complier-average causal effect (CACE).

This estimand can be estimated via the ratio estimator or
 the two-stage least squares (TSLS) approach \cite{angrist1995two}.  The later consists of a ``first stage'', which regresses treatment received on randomised treatment, and a ``second stage'',  which models the outcome on the predicted treatment received.
 Additional covariates can be included in each stage to  control for measured confounding or increase precision.
 The regression coefficient for the predicted treatment received in the second stage model is a consistent estimator of the LATE, provided that the first stage model is a linear regression, containing all the variables appearing in the second stage \cite{Robins2000, Wooldridge2010}.

Extensions of this approach for the estimation of LATE in CRTs have been proposed, ranging from a TSLS of individual-level data with variance inflation by the design effect factor \cite{angrist2008mostly}, to  multilevel mixture models that include the latent compliance class membership as a regressor and a random effect for cluster \cite{frangakis2002clustered, jo2008cluster}. Schochet \cite{Schochet2013} uses cluster-level (CL) summaries for both treatment received and outcome to construct a ratio estimator. 

In this paper we focus on TSLS estimation applied to CL outcome summaries. Similar to \cite{Schochet2013}, this approach exploits well-known methods from cluster-level analysis, which consist of calculating for each cluster a relevant summary measure of the individual-level outcomes, such as means or proportions, and then analysing these using  appropriate statistical methods, such as regression.  Because each cluster  provides only one data point, the units of analysis can be considered to be independent,  but the procedure is inefficient \cite{campbell2000analysis}.  Estimation by weighted least squares, where the weights  are defined either by the cluster size or by the so-called minimum variance weights can improve efficiency \cite{angrist2008mostly}. 
 Comparing these alternative estimation strategies for the implementation of TSLS estimation using CL data is the focus of this paper.

 The rest of the paper proceeds as follows. Section  \ref{Sec:Methods} provides an overview of cluster level analysis methods, defines the estimand of interest, the LATE, and introduces the identification assumptions  and the different cluster-level TSLS approaches.
Section \ref{Sec:Simulation} presents the simulations performed to compare the finite-sample performance of the  methods considered. In Section \ref{Sec:TXT} we illustrate the methods by re-analysing the TXT4FLUJAB trial, a UK based CRT evaluating the effectiveness and efficacy of text messaging influenza vaccine reminders in increasing vaccine uptake amongst patients with chronic conditions \cite{herrett2016text}.  Section \ref{Sec:Discussion} concludes with a discussion.

\section{Methodology} \label{Sec:Methods}

Consider  a  two-arm CRT, with $n$  participants, indexed by $i$,  in $J$ clusters, indexed by $j$, each of size $n_j$. Let $Z_j$ denote the binary treatment randomly allocated  at the cluster-level with probability 0.50.
Let $Y_{ij}$ denote the continuous or binary outcome, and $D_{ij} \in \{0,1\}$ be the treatment received by individual $i$ in cluster $j$. Let $W_j$ and $X_{ij}$ be baseline covariates at cluster and individual-level, respectively (which can be vectors of variables).

With a slight abuse of notation, we let $Y_j$ denote the CL  outcome (mean or proportions), {\it i.e.} $Y_j\!=\!\frac{1}{n_j} \sum\limits_{i=1}^{n_j} Y_{ij}$,  hereafter referred to  as the unadjusted CL outcome. Analogously, let $D_j$ denote the unadjusted CL treatment received,
 $D_j\!=\!\frac{1}{n_j} \sum\limits_{i=1}^{n_j} D_{ij}$.

In the cluster-level adherence settings, $D_{ij}$ is constant within clusters, and therefore  $D_j$ is binary.  In contrast, when non-adherence is at the individual level, $D_j$ is a continuous measure that varies from 0 to 1, representing  the proportion of individuals receiving the active treatment in cluster $j$.

\subsection{Cluster-level analysis}
The unadjusted CL analysis, uses simple CL summary statistics as the outcomes in subsequent  analyses.
Let $\sigma^2$ denote the variance of $Y_{ij}$, which can be decomposed as $\sigma^2 = \sigma_{\epsilon}^2 + \sigma_{\upsilon}^2$, where  $\sigma_{\upsilon}^2$ is the between-cluster variance and $\sigma_{\epsilon}^2$ the within-cluster variance.  The intra cluster correlation coefficient (ICC) for $Y_{ij}$ is then $\rho_y= \frac{\sigma_{\upsilon}^2}{\sigma_{\upsilon}^2 \!+ \!\sigma_{\epsilon}^2}$.
  The variance of $Y_j$ is
\begin{equation}
\label{VayY}
\mbox{Var}(Y_j) \!=\! \sigma_{\upsilon}^2 \!+\! \frac{\sigma_{\epsilon}^2}{n_j} \!=\! \dfrac{1 \!+\! \rho_y(n_j \!-\! 1)}{n_j}\sigma^2.
\end{equation}

 Since CL outcomes are continuous (regardless of whether the original variable was binary), they can be thought to be approximately normally distributed  (if $n_j$ is sufficiently large). Thus a linear regression with  CL outcome $Y_j$ as dependent variable and $Z_j$ as the explanatory variable can be fitted to estimate the ITT effects. In the simplest setting, we have
 \begin{equation}\label{simplereg}
Y_j=\alpha_0 + \alpha_Z Z_j +\eta_j
\end{equation}
where $\eta_j$ is a random error term, assumed to be independently and identically distributed (i.i.d.), with mean 0. The ITT is estimated by $\alpha_Z$.

Efficiency is gained by estimating this model using generalised least squares (GLS), with the weights  being either the cluster size $n_j$, or  the so-called \emph{minimum-variance (MV) weights} given by \cite{Prais1954},
$$\omega_j = \frac{n_j}{1+\rho_y(n_j-1)}.$$

MV weights  are approximately equivalent to cluster size weights  when $\rho_y \approx 0$, while if $\rho_y \approx 1$, MV weights are approximately 1 \cite{Campbell2014}. These equivalences can have practical implications when  the variance of  $\eta_j$ cannot be consistently estimated, for example  if the number of clusters is small, so weighting by the cluster size  or even no weights, are viable alternatives. Where clusters are large, weighting by  cluster size is inefficient \cite{kerry2001unequal}.

Since the  $\eta_j$ can be heteroscedastic especially when cluster sizes are very imbalanced, the standard errors should be obtained using a method that takes this into account, such as the  Huber-White standard errors (HW SE) \cite{white1980heteroskedasticity} which are consistent when there is heteroscedasticity \cite{cameron2005microeconometrics}.

Finally, because each cluster now contributes only one observation, inference should be based on the number of clusters $J$. Therefore, if $p$ is the number of parameters being  estimated, hypothesis tests and confidence intervals (CIs) should be based on appropriate distributions, for example $t_{J-p}$ and not on normal-based approximations. We refer to this as small-sample degrees of freedom (SSDF) correction.  Where $J$ is sufficiently large ($>40$) normal approximations are adequate.

Regression analyses of CL summary outcomes can only adjust for CL covariates directly, as  using CL summaries for individual-level regressors is not appropriate \cite{angrist2008mostly}. However, where there is interest in adjusting for baseline covariates at the individual level this can be done  through a two-step procedure \cite{hayes2009cluster}.  First, an individual-level regression analysis of the outcome  is performed incorporating all the relevant covariates into the regression model except for the treatment indicator and ignoring clustering , e.g. with only one covariate $X_{ij}$, we have:
\begin{equation}\label{adjreg}
Y_{ij}= \lambda_0+\lambda_1 X_{ij} +e_{1_{ij}}.
\end{equation}

In the second step, the sample mean of the fitted  residuals for this model $\widehat{e}_{1_{ij}}$  is calculated for each cluster $j$, $$e_j = \frac{1}{n_j} \sum_{j=1}^{n_j} \widehat{e}_{1_{ij}}.$$
These are then used as  CL outcomes in any subsequent analyses. See the Appendix for the formulation for binary outcomes.

We refer to these summaries as adjusted CL  outcomes (adCL). Regression models involving them can  also be estimated by GLS, with inference based on normal approximations or HW SEs  and/or SSDF corrections, as before.
Of note,  if CL covariates are used to compute adCL outcomes,  the degrees of freedom (DF) must be further reduced by the number of cluster-level regressors used to obtain the CL outcome. No such adjustment is necessary for individual level variables \cite{hayes2009cluster}. In this work, we only use adCL outcomes obtained by adjusting for individual level variables. In the remainder, we  denote the CL summary outcomes by $Y_j$, whether they are unCL or adCL, will be clear  from the context.

\subsection{LATE for CL data}\label{Sec:Ident}

\subsubsection{Notation and technical assumptions}\label{Sec:IVassumptions}

Denote by $Y_{ij}(\mathbf{d}_j)$ the \emph{potential outcome} that would manifest if, possibly contrary to fact, the $j$-th cluster to which the individual belongs receives treatment  $\mathbf{d_j}$, a vector of length $n_j$ of 0s and 1s, where we are assuming \textit{no interference between clusters}, i.e. the potential outcomes and potential treatment received of individuals in the $j$-th cluster are unrelated to the treatment status of individuals in other clusters  \cite{schochet2011estimation}.  ``No interference between clusters'' is a special case of partial interference,  where individuals can be partitioned into groups such that interference does not occur between individuals in different groups but may occur between individuals in the same group \cite{Sobel2006}. This is commonly assumed in clustered randomised trials \cite{frangakis2002clustered,jo2008cluster}.
We also assume \textit{conterfactual consistency}: for  $j=1,\ldots, J$, if $Z_j=z$ then $D_{ij} = D_{ij}(z) \mbox{ and } Y_{ij} = Y_{ij}(z,  D_{ij}(z))$, for all $i=1,\ldots,n_j$.

\subsubsection{Estimand of interest and identification assumptions}
Assuming no interference between clusters and consistency, allows us to define the estimand of interest, the local average treatment effect \cite{imbens1994identification}.  

In the setting considered here where both $Z_j$ and $D_{ij}$ are binary, the vector of potential treatment received under alternative random allocation, $(D_{ij}(0), D_{ij}(1))$ partitions the participants in each cluster  into four different \textit{compliance classes}: $C_{ij} = n$ (never-takers) if $D_{ij}(0) = D_{ij}(1) = 0$; $C_{ij} = a$ (always-takers) if $D_{ij}(0)= D_{ij}(1) = 1$; $C_{ij} = c$ (compliers) if $D_{ij}(z) = z$ for $z\in\{0,1\}$; and $C_{ij} = d$ (defiers) if $D_{ij}(z) = 1-z$ for $z\in\{0,1\}$.

The estimand  of interest here is the so-called  \emph{population} complier average causal effect (LATE), defined as
\begin{eqnarray}
\nonumber
\beta&= &E_jE_i\left[\{ Y_{ij}(1,  D_{ij}(1))- Y_{ij}(0,  D_{ij}(0))\}\vert  C_{ij} = c \right]\\
&= &\frac{\sum_{j=1}^J\sum_{i=1}^{n_j} \{Y_{ij}(1,  D_{ij}(1))- Y_{ij}(0,  D_{ij}(0))\}\{ I(D_{ij}(1) = 1,D_{ij}(0) = 0)\}}
{\sum_{j=1}^J\sum_{i=1}^{n_j}I(D_{ij}(1) = 1,D_{ij}(0) = 0)}
\end{eqnarray}
This is said to be a ``local'' causal effect as it is conditional on the stratum of complier individuals. For this reason, this is often referred to as the \emph{local average treatment effect} (LATE).

Following \cite{schochet2011estimation, Kang2018}, we write the cluster version of the corresponding identification assumptions  \cite{angrist1996identification}  as follows:\\

\noindent\textbf{(A1)}  \textbf{Cluster unconfoundedness} : $Z_j \ci  D_{ij}(z),Y_{ij}(z,D_{ij}(z)),\quad z\in\{0,1\}.$ This is also known as cluster or group  independence or cluster randomisation assumption.
\\
\noindent\textbf{(A2)}  \textbf{Exclusion restriction at the individual level} : Conditional on the treatment received $D_{ij}=d$, the treatment assignment $Z_j$ had no effect on the outcome. In terms of potential outcomes we have:

\begin{equation*}
Y_{ij}\left(1,d\right) = Y_{ij}\left(0,d\right)\quad \forall d\in\{0,1\}.
	\end{equation*}
\noindent\textbf{(A3)} \textbf{Instrument relevance:} Also referred to as first stage assumption:

$Z_j$ is causally associated with treatment received $D_{ij}$, i.e.
 $Z_j \not\perp\!\!\!\perp D_{ij}$.

For point identification of local treatment effects,  \textbf{(A4) monotonicity} of the treatment mechanism is often assumed:  $D_{ij}(1) \geq D_{ij}(0)$ , often informally referred to as ``there is no defiers'' \cite{imbens1994identification}. Notice we need to assume this holds at the individual level \cite{Kang2018}. For the cluster-level non-adherence setting, where $D_{ij}$ does not vary within clusters, then this becomes monotonicity at the cluster level $D_{j}(1)=1, \ D_{j}(0)=0$. 

An extra assumption necessary when  using adjusted CL outcomes is that the  model used to derive them is correctly specified.

\subsubsection{Cluster and individual-level non-adherence}

 The population-level LATE estimand $\beta$ can be thought of as a weighted average of the cluster-specific LATE for each cluster $j$, $\beta_j$, namely
\begin{equation} \label{popLATE}
  \beta=\sum_{j=1}^{J} \psi_{j} \beta_j,
\end{equation}
where
\begin{eqnarray}
\nonumber
\beta_{j}&=&
E_i\left[ Y_{ij}(1,D_{ij}(1))- Y_{ij}(0,D_{ij}(0)) \vert C_{ij} = c \right]\\
&=&\frac{1}{n_{c,j}}\sum_{i=1}^{n_j}\left[ \{Y_{ij}(1,D_{ij}(1))- Y_{ij}(0,D_{ij}(0))\}\{I(D_{ij}(1)=1,D_{ij}(0)=0)\}\right],
\end{eqnarray}
with $n_{c,j}$ is the number of individual-level compliers in each cluster $j$,  assumed here to be $>0$ for all clusters.
The weights corresponding to each $\beta_j$  are
$\psi_{j}= \frac{n_{c,j}}{\sum_{j=1} ^J n_{c,j}}$,
that is the number of cluster-specific compliers divided by the total number of compliers.

This result is useful when interpreting the estimates obtained using CL summaries. We first note that the Wald estimand applied to CL summaries, $\beta_{CL}$ does not always correspond to the population LATE $\beta$.  The former can be expressed as \cite{schochet2011estimation}:

\begin{equation} \label{LATEformula}
\beta_{CL}  = \frac{E[Y_j\vert Z_j=1] - E[Y_j\vert Z_j=0]}{E[D_j\vert Z_j=1] - E[D_j\vert Z_j=0]}
\end{equation}
In the case where  treatment received is at the cluster level (\textit{i.e.} cluster-level adherence), this  CL Wald estimand  indeed can be interpreted as a the  population LATE.

In the case where non-adherence varies at the individual level, it can be shown that
$\beta_{CL}  =\sum_{j=1}^J {\psi_{CL,j}}\beta_{j}$
where the CL-weights are
${\psi_{CL,j}}= \frac{n_{c,j}/n_j}{\sum_j n_{c,j}/n_j}$, i.e. the normalised  proportion of individual compliers in each cluster \cite{Kang2018}.
 So, CL-LATE $\beta_{CL}$ identifies the population LATE , equation \eqref{popLATE}, only if (i) the cluster sizes $n_j$ are identical for all $j$, or (ii) the cluster-specific LATE $\beta_j$  are homogeneous across all clusters.

In the remainder, with individual-level non-adherence we assume that every cluster has the same cluster-specific LATE, but allow for the cluster sizes to vary. If this is not the case,  the CL-LATE $\beta_{CL}$ identifies a weighted average of the heterogeneous cluster-specific LATE, because of clusters with the same proportions of compliers are weighted the same, without accounting for the cluster size.

\subsection{TSLS for CL data}\label{Sec:TSLS}

The conditional expectations appearing in the Wald estimand (equation  \ref{LATEformula}) can be estimated via TSLS regression of the CL summaries (referred to as CL-TSLS). CL-TSLS is most easily explained for  settings without weights or covariate adjustment.
The first stage fits  a regression to CL treatment received $D_j$ on treatment assigned $Z_j$. Then, in a second stage,  a regression for the  CL outcome, (either unCL or adCL),  on the predicted treatment received is fitted. Crucially both first and second stages must be linear models for the TSLS estimator to be guaranteed to be consistent \cite{Wooldridge2010, Vansteelandt2018}. We have:
\begin{eqnarray}\label{2sls}
\nonumber D_{ j} &= &\gamma_0 + \gamma_{ Z}Z_{ j} + \omega_{1{ j}}\\
Y_{j} &=& \beta_0 + \beta_{IV} {\widehat{D_{ j}}} + \omega_{2{j}}
\end{eqnarray}
\noindent where $\omega_{1j}$ and $\omega_{2j}$ are assumed i.i.d. with mean zero and constant variance, with  $\omega_{1j}\ci\omega_{2j}$, and 
${\widehat{\beta}}_{IV} $ is the estimate of CL-LATE.

Cluster-level covariates can be included to increase precision. For example, with one CL covariate $W_j$, we have
\begin{eqnarray}\label{2slsw}
\nonumber D_j &=& \gamma_0 + \gamma_Z Z_j + \gamma_W W_j + {\upsilon_{1j}} \\
Y_j &=& \beta_0 + \beta_{IV} {\widehat{D}}_j + \beta_W W_j +  {\upsilon_{2j}}
	\end{eqnarray}
where the error terms are as before. 

The asymptotic standard error for the TSLS estimator is given in \cite{imbens1994identification}, and implemented in commonly used software packages. 

As with the cluster level estimation of ITT, where the number of clusters is small,  the CIs are constructed using a $t$ distribution with degrees of freedom equal to $J-p$, where $p$ is the number of parameters estimated by the second stage (i.e. the SSDF correction). 
As before,  MV  or cluster size weights can be used to increase efficiency.
Finally, the error terms in the CL-TSLS are assumed to be homoscedastic.  Where this is not a sensible assumption, Huber-White SEs should be used \cite{white1982instrumental}.

\section{Simulation study}\label{Sec:Simulation}

We now perform  a simulation study comparing the finite sample performance of  TSLS estimation applied to CL data. 
We simulate CRT individual-level data with one-way non-adherence,  at either cluster or  individual level. 
With a fixed expected total sample size equal to $1000$, we vary the number of clusters $J$, and the average cluster size $n_j$. The marginal ICC of $Y$ also takes two values.  The effect of individual and cluster level variables on the outcome and the treatment received also varies, so that the strength of the confounding is either low or high, while  the value of the true LATE also has two levels. Table \ref{Scenario} summarises the factorial design and the values taken by the different levels.

 More specifically, we  simulate cluster randomised treatment $Z_j\sim \mbox{Bern}(0.5)$ and two independent baseline covariates, a cluster-level covariate $W_j \sim N(0,\sigma^2_W)$ and individual-level covariate $X_{ij}\sim N(0,\sigma^2_X)$ with a moderate ICC $\rho_X=0.05$, and $\sigma^2_W=\sigma^2_X=0.08$.

We then generate a binary adherence class indicator variable $C_{ij}$, which is considered as latent. For settings where adherence is at the cluster level,  this is constant within clusters, under the following model
\begin{eqnarray*}
C_{ij}&=&C_j \sim \mbox{Bern}(\pi_j)  \quad \mbox{with} \quad \pi_j = P(C_j \!=\! 1) \\
\mbox{logit}(\pi_j) &=& \lambda_0 + \lambda_W W_j,
\end{eqnarray*}
with $\lambda_W \!=\! 0.05$ equivalent to an odds ratio OR $\approx 1.05$ per unit increase in $W$ (denoted ``small effect'') and $\lambda_W \!=\! 0.7$ equivalent to OR $\approx 2$ (``large effect'').

For settings with individual-level adherence, the data generating model is
\begin{align*}
C_{ij} \sim \mbox{Bern}(\pi_{ij}) & \quad \mbox{with} \quad \pi_{ij} = \pi = P(C_{ij} = 1) \\
\mbox{logit}(\pi_{ij}) & = \lambda_0 + \lambda_W W_j + \lambda_X X_{ij} + \zeta_j \\
\zeta_j & \sim N\Big(0,\sigma^2_{\zeta}\Big)
\end{align*}
with $\sigma^2_{\zeta} \!=\! \pi^2/3$, so that  the ICC for compliance is $\rho_C \!=\! \sigma^2_{\zeta}/(\sigma^2_{\zeta} + \pi^2/3) \!=\! 0.50$.

We derive treatment received at the individual level as
$$D_{ij}=Z_jC_{ij},$$
so that those individuals in clusters randomly allocated to control have always control treatment, but those in clusters randomised to the active intervention can switch to the control treatment, depending on their adherence class.

We finally generate continuous outcome $Y_{ij}$, under the exclusion restriction assumption, 
 \begin{equation} \label{simYij_ClusAdh}
\begin{aligned}
Y_{ij} &=\beta_0 + \beta_C C_{ij} + \beta_{CZ} C_{ij} Z_j + \beta_W W_j + \beta_X X_{ij} + \upsilon_j + \epsilon_{ij}
\end{aligned}
\end{equation}
with $\upsilon_j \sim N(0,\sigma^2_{\upsilon})$ and  $\epsilon_{ij} \sim N(0,\sigma^2_{\epsilon})$, where the values for  $\sigma^2_{\upsilon}$ and $\sigma^2_{\epsilon}$ are chosen such that the  marginal ICC for $Y$ has the corresponding value according to the simulated scenario, given that $\mbox{Var}(Y_{ij}) = \sigma^2=1$. 
The choice of the parameters' values is reported in Table \ref{Scenario}.

\setlength{\LTleft}{0pt}
{\fontsize{10}{11.4}\selectfont
	\begin{longtable}{llll}
		\caption{Factorial design of the data generating processes and values taken by the parameters in the simulations.} \label{Scenario} \\
		\hline
		\noalign{\vskip 1mm}
		\multicolumn{1}{l}{{\textbf{Parameter}}} & \multicolumn{1}{l}{{\textbf{Label}}} & \multicolumn{1}{l}{{\textbf{Level}}} 	 & \multicolumn{1}{l}{{\textbf{Value}}} \\
		\endfirsthead
		\caption*{\textbf{Table 2 Continued}} \\
		\hline
		\multicolumn{1}{l}{{\textbf{Parameter}}} & \multicolumn{1}{l}{{\textbf{Label}}} & \multicolumn{1}{l}{{\textbf{Level}}} 	 & \multicolumn{1}{l}{{\textbf{Value}}} \\
		\midrule
		\endhead
		\bottomrule
		\multicolumn{4}{r@{}}{Continued on next page} \\
		\endfoot
		\endlastfoot
		\midrule
		
		\multicolumn{4}{l}{\textbf{\emph{CRT size}}}  \\
		\hspace{1em} $n$ & Total number of individuals & Moderate            	& $\approx 1\;000$ \\
		\hspace{1em} $J$ & Number of clusters  and  & Moderate clusters 	& $J=50, n_j \sim \text{Poi}(20)$ \\
	\hspace{1em}$n_j$& individuals per cluster  & Few large clusters & $J=10, n_j \sim \text{Poi}(100)$ \\
		\noalign{\vskip 2mm}
		
		\multicolumn{4}{l}{\textbf{\emph{Baseline variables}}}  \\
		\hspace{1em} $W_j$     & Cluster-level variable    & -         	& $W_j \sim N(0,0.08)$ \\
		\hspace{1em} $\rho_{X}$& ICC for $X_{ij}$ & Moderate & 0.05 \\
		\hspace{1em} $X_{ij}$  & Individual-level variable & -     		& $X_{ij} = X_j + e_{ij}, X_j \sim N(0,0.004)$, \\
		\hspace{1em}  &  & & $e_{ij} \sim N(0, 0.076)$ \\
		\noalign{\vskip 2mm}
		
		\multicolumn{4}{l}{\textbf{\emph{Adherence to treatment}}} \\
		\hspace{1em} $\pi$ & Expected probability & Moderate & 0.60 (cluster-level adherence) \\
		& of adherence & & 0.85 (individual-level adherence) \\
		\hspace{1em} $\lambda_W$, $\lambda_X$ 	& $W_j$ and $X_{ij}$ effects on & Small & $\lambda_W=0.05, \lambda_X=0.05$ \\
		\hspace{1em} & log odds of adherence 	& Large & $\lambda_W=0.70, \lambda_X=0.70$ \\				
		\hspace{1em} $C_j$  & Cluster-level adherence 	& -  & Bern[$\texttt{expit}(\lambda_0 + \lambda_W W_j)]$ \\
		                    & class     &   & \\
		\hspace{1em} $C_{ij}$  & Individual-level    & -  & Bern[$\texttt{expit}(\lambda_0 + \lambda_W W_j + $ \\
		\hspace{1em} & adherence class  & & $\qquad \qquad \qquad \lambda_X X_{ij} + \zeta_j$)] \\
		\hspace{1em} $\zeta_j$  & Cluster-level random effects  & -  & $\zeta_j \sim N(0,\pi^2/3)$ \\
		\hspace{1em} $\rho_{C}$& ICC for $C_{ij}$ & Moderate & 0.50 \\
		\noalign{\vskip 2mm}
		
		\multicolumn{4}{l}{\textbf{\emph{Outcome}}} \\
		\hspace{1em} $\beta_0$, $\beta_C$  &            &       & $\beta_0$=0, $\beta_C$=0 \\						
		\hspace{1em} $\beta_W$, $\beta_X$  & $W_j$ and $X_{ij}$ effects & Small & $\beta_W$=0.1 SD, $\beta_X$=0.1 SD \\				
		\hspace{1em} & on outcome $Y_{ij}$ & Large & $\beta_W$=0.4 SD, $\beta_X$=0.4 SD \\
		\hspace{1em} $\beta_{CZ}$ & True LATE & Small, Large & 0.1 SD, 0.4 SD \\
		\hspace{1em} $\rho_Y$  & ICC for $Y_{ij}$ & Small, Large & 0.05, 0.20 \\	
		\bottomrule
		\multicolumn{4}{l}{$^\text{a}$ \scriptsize{SD: standard deviation of the outcome $Y$,  $\sigma= 1$.}}
	\end{longtable}
}

We need  the data generating process to result in randomised treatment $Z$ being a valid IV, but some combinations may result in weak instruments, for example, cluster-level non-adherence settings, with only 5 clusters per arm, and the proportion of non-adherent clusters set  at 40\%  (the median proportion of non-adherent clusters  reported in  \cite{agbla2017adherence} being 44.8\%).
Thus, after creating each dataset, we perform an unadjusted first stage regression of $D_j$ on  $Z_j$  and reject simulated datasets where  the resulting $F-$statistic is $<10$,   (Staiger \& Stock's rule of thumb  for weak instruments \cite{staiger1994instrumental}).
We continue this process until we have $2500$ datasets per scenario.

Estimation  in each scenario involves using unadjusted CL summary of treatment received in the first-stage, and either unadjusted or individual-level variable adjusted CL summary outcomes, for the second stage. Each regression in the TSLS was fitted via OLS or GLS, the latter with either  cluster size or MV weights. We also consider TSLS  where each stage model is either unadjusted or adjusted for a cluster-level variable.
Finally, we obtain SEs assuming homoscedasticity or HW SEs,  and SSDF-based  or normal approximation CIs.  An summary is given in Table \ref{Overview}.
Details of the Stata code used for analysis are found in the web-Appendix.

	\begin{table}[h]
	\caption{Overview of TSLS estimation and inference strategies used in the simulation study}
    \label{Overview}
    	\begin{tabular}{llll}
		\hline
        \textbf{Analyses features}&\multicolumn{3}{c}{{\textbf{Levels}}}  \\
        \hline
        CL outcome  &  Unadjusted & Adjusted for $X_{ij}$ \\
        TSLS adjusted for $W_j$ : &  No& Yes\\
        Weights  & none (i.e. OLS) & CS  &  MV \\
        SE estimation:&    Normal theory &   HW SE \\
        SSDF correction:&  No &  Yes\\
        \hline
        \end{tabular}
        \\ \\
        {\footnotesize CL: cluster level; HW: Huber-White; CS weights: cluster-size weights; MVW: minimum variance weights; SE: standard error; SSDF: Small sample degrees of freedom correction}
   \end{table}

The performance criteria used are empirical bias and coverage rates of the 95\% CIs over the $2500$ replicate datasets per scenario. For the bias,  we construct a 95\% CI  using its Monte Carlo Error (MCE). The coverage rate sampling error given the size of the simulation results in a valid range between  94.1\% and 95.9\%. See the Appendix for the formal definitions.

\subsection{Results}

We present the results by plotting the empirical bias  with the MCE-based CIs. The coverage rate valid range is represented by horizontal dashed lines. 

\begin{figure}
\thispagestyle{empty}
		\begin{center}
		\includegraphics[scale=0.5]{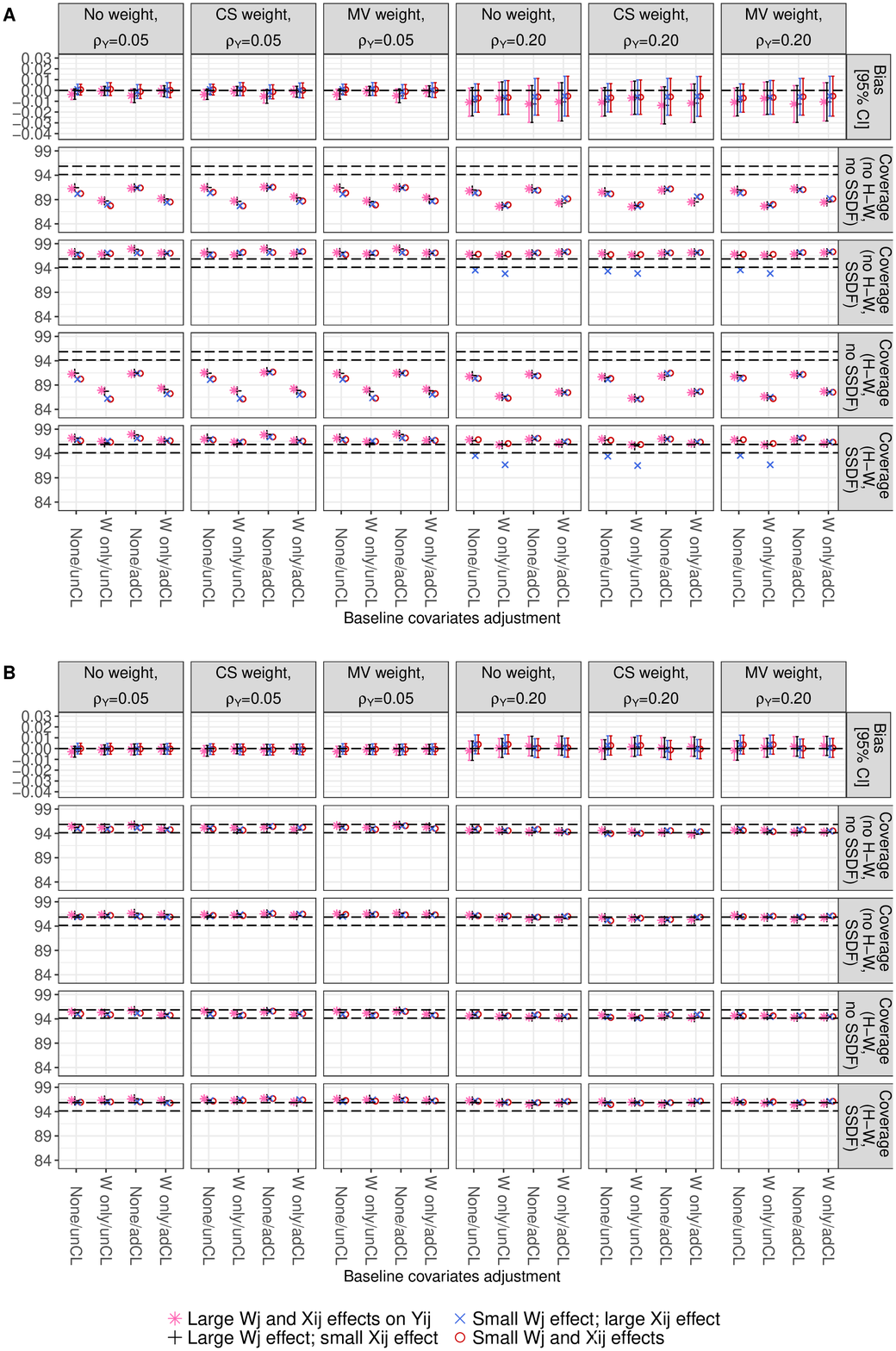}
		\caption{\small{Bias (top row) and 95\% CI coverage (rows 2--5) of CL-LATE with cluster-level non-adherence and  large true LATE.
Data generation scenarios represented by $\ast, +, \times$, and $\circ$. Estimates are obtained via unadjusted or $W$-adjusted TSLS with different weights (none, cluster size (CS) and minimum-variance (MV)) (by column) using  CL unadjusted or adjusted for $X$ outcomes (``unCL'' or ``adCL''). Small  ($J=10$) and large ($J=50$) number of clusters results are shown in Panel A  and B .}} 		\label{fig:CluAdh_LargeCACE_Ybar_eligible}
   \end{center}
  	\end{figure}

\begin{figure}
\thispagestyle{empty}
	\centering
		\includegraphics[scale=0.5]{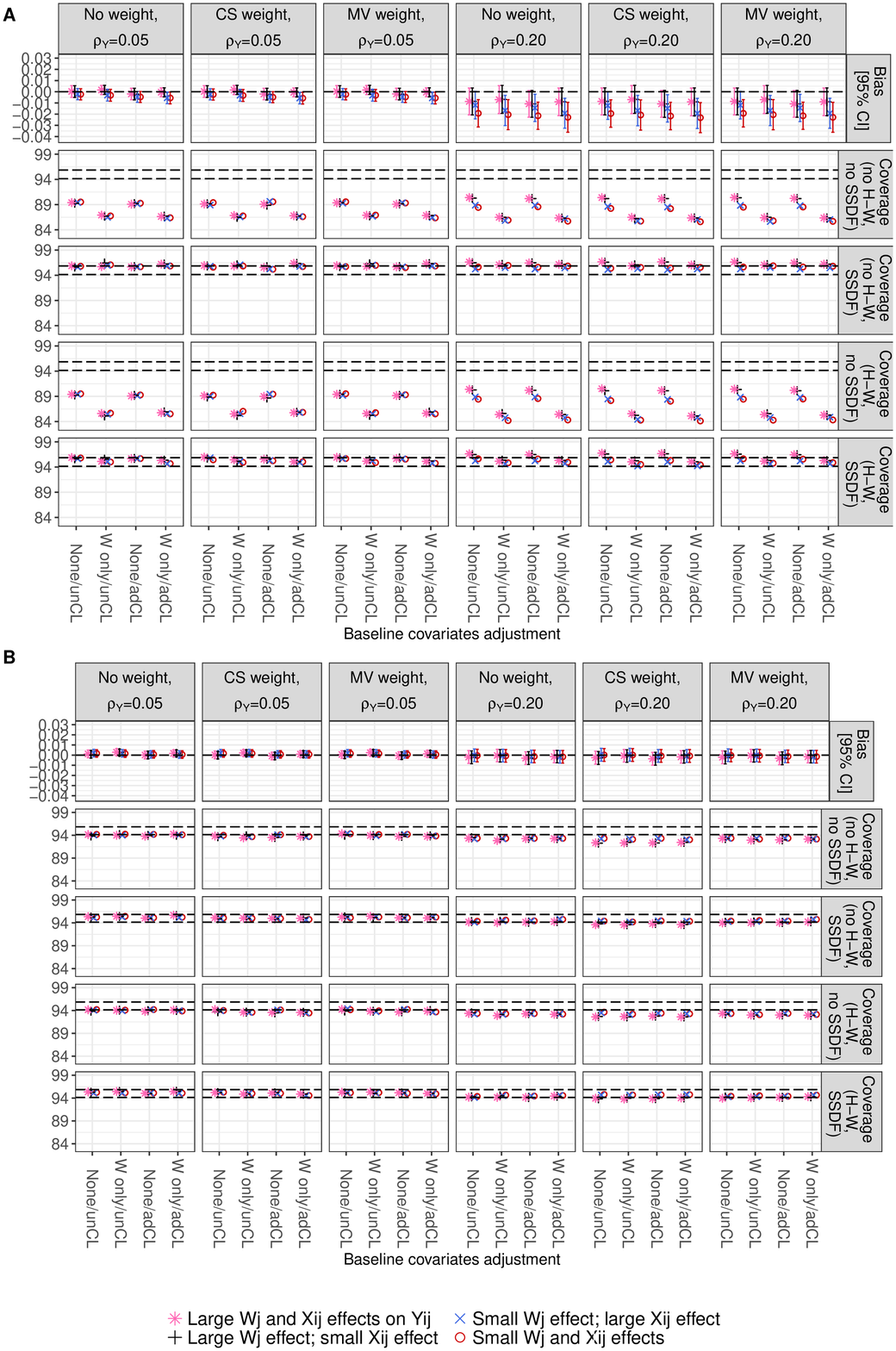}
			\caption{\small{Bias (top row) and 95\% CI coverage (rows 2--5) of CL-LATE with individual-level non-adherence and  large true LATE. Data generation scenarios represented by $\ast, +, \times$, and $\circ$. Estimates are obtained via unadjusted or $W$-adjusted TSLS with different weights (none, cluster size (CS) and minimum-variance (MV)) (by column) using  CL unadjusted or adjusted for $X$ outcomes (``unCL'' or ``adCL''). Small  ($J=10$) and large ($J=50$) number of clusters results are shown in Panel A  and B .}}
		\label{fig:IndAdh_LargeCACE_Ybar_eligible}
	\end{figure}

\begin{figure}
\thispagestyle{empty}
		\centering
		\includegraphics[scale=0.5]{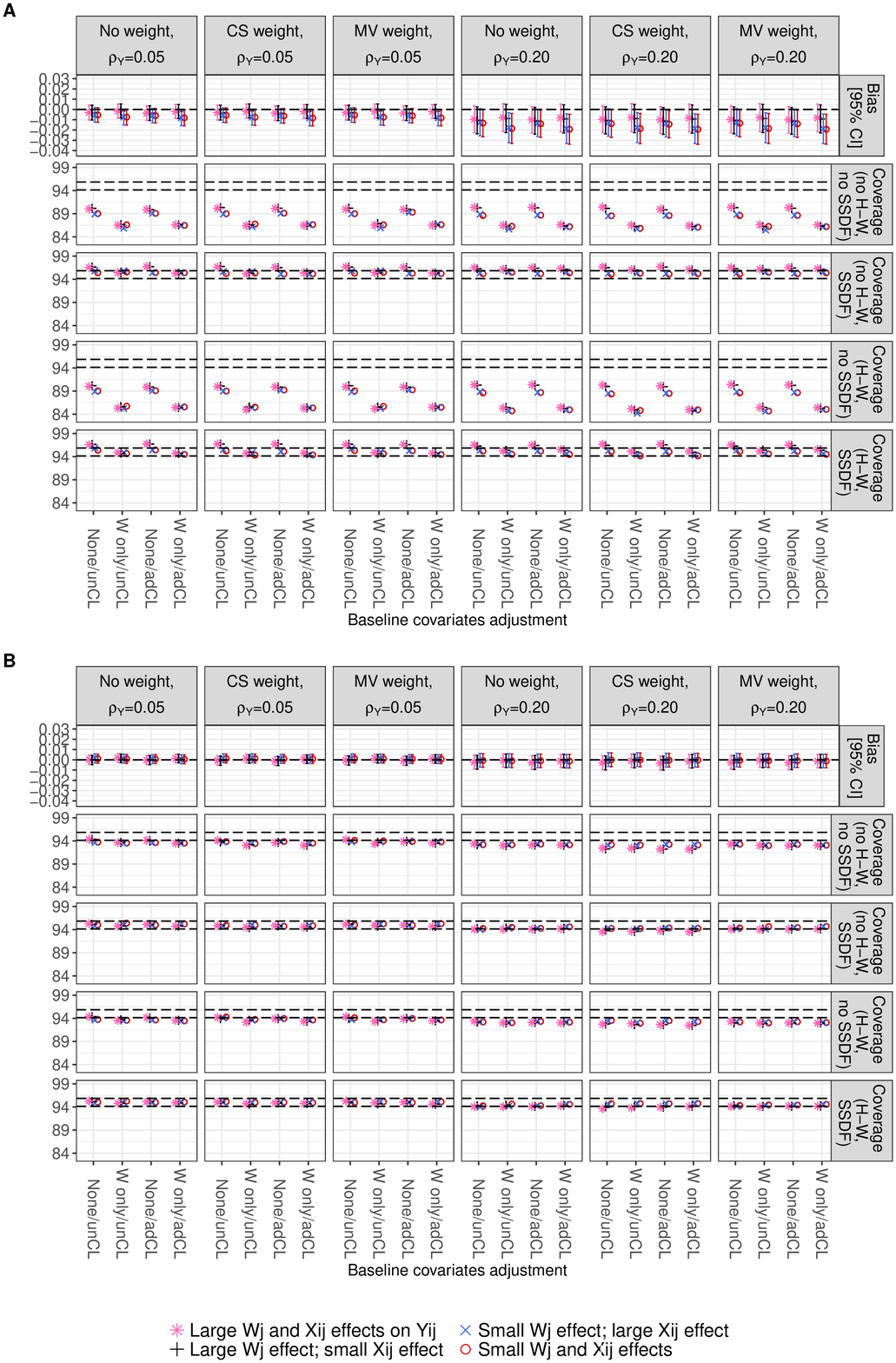}
\caption{\small{Bias (top row) and 95\% CI coverage (rows 2--5) of CL-LATE with individual-level non-adherence and  small true LATE. Data generation scenarios represented by $\ast, +, \times$, and $\circ$. Estimates are obtained via unadjusted or $W$-adjusted TSLS with different weights (none, cluster size (CS) and minimum-variance (MV)) (by column) using  CL unadjusted or adjusted for $X$ outcomes (``unCL'' or ``adCL''). Small  ($J=10$) and large ($J=50$) number of clusters results are shown in Panel A  and B .}}		

				\label{fig:IndAdh_SmallCACE_Ybar_eligible}
	\end{figure}

Figure \ref{fig:CluAdh_LargeCACE_Ybar_eligible} and Figure \ref{fig:IndAdh_LargeCACE_Ybar_eligible} report the empirical bias and 95\% CI coverage resulting from each of the different CL-TSLS estimators, when adherence is at cluster  or individual level respectively, and for scenarios where the true  LATE is large (0.4 SD).
The corresponding  figures for small true LATE are in the Appendix, Figure \ref{fig:CluAdh_SmallCACE_Ybar_eligible} and \ref{fig:IndAdh_SmallCACE_Ybar_eligible}.

Each figure reports results where $J=10$ (Panel A, top) or $J=50$ (Panel B), and with the ICC for $Y$, $\rho_Y$, is either small (first three columns) or large (last three columns). In each cell,  the results for alternative combinations of TSLS (unadjusted/adjusted for $W_j$) applied to unCL or adCL outcomes are plotted along the horizontal axis.The different data generation scenarios are identified by $\ast, +, \times$, and $\circ$, corresponding to varying strengths of the effects of $X$ and $W$ on $Y$.    

We see that all CL-TSLS estimators show finite sample bias, regardless of whether the non-adherence was at the cluster or individual level and whether the CL summary for $Y$ was adjusted or unadjusted, or $W_j$ was included or not in  the TSLS regressions. The bias  is more severe when the ICC for $Y$ is larger (right hand side of each Figure), especially if the number of clusters is small (Panel A). The bias is somewhat attenuated when  we adjust for $W_j$ in the TSLS, and the non-adherence is at the cluster-level (Figures \ref{fig:CluAdh_LargeCACE_Ybar_eligible} and \ref{fig:CluAdh_SmallCACE_Ybar_eligible}).
 In contrast, for settings with individual-level non-adherence,  this adjustment instead increases the bias, especially if $W$ has only a small confounding effect. In these scenarios, the estimates exhibit a small but statistically significant bias, which disappears when the number of clusters is larger (Figures \ref{fig:IndAdh_LargeCACE_Ybar_eligible} and \ref{fig:IndAdh_SmallCACE_Ybar_eligible}).
 In general, the bias is not affected by the choice of weighting strategy, nor by whether $\rho_Y$ is small or large.

Comparing the results of the 2$^{nd}$,  3$^{rd}$, and 4$^{th}$ rows in each panel  (Figures \ref{fig:CluAdh_LargeCACE_Ybar_eligible} and \ref{fig:IndAdh_LargeCACE_Ybar_eligible}),  we see that the coverage rate is affected by the choice of SE estimation and also by whether SSDF correction is used.
When the number of clusters is small, an SSDF  correction must be used as  failing to do so results in under-coverage (Panels A). The low coverage is more serious when TSLS adjusts for $W$ (second and fourth set of results in each panel).

Overall, the results in Panel A of each figure show that using HW SE or not has little to no impact if there is no SSDF correction.  However, when the SSDF correction is used for settings with  cluster-level non-adherence, large $\rho_Y$, and large true LATE, but where only $X$ is strongly associated with $Y$, using unCL outcomes leads to under-coverage, regardless of weighting or SE method (Figures \ref{fig:CluAdh_LargeCACE_Ybar_eligible} and \ref{fig:CluAdh_SmallCACE_Ybar_eligible}, 3$^{rd}$ and 5$^{th}$ rows of Panel A, right hand side columns). The use of adCL outcomes (\textit{i.e.} where the CL outcome is the residual after adjusting for individual level variable $X$) recovers coverage close to nominal.
 This is not the case  when the non-adherence is at the individual level, and both $W$ and $X$ are confounders of the causal effect the treatment received $D$ and the outcome $Y$ in the data generating process.

 In both cluster and individual-level non-adherence settings, it can be seen that  using MV weights increases the coverage by a small fraction, when compared with cluster size weights, especially for scenarios with  $J=50$ and large $\rho_Y$. However, since MV weights require an estimate of the cluster-level variance, and this is badly estimated when the number of clusters is small ($J=10$), we can see that MV weights are less efficient than using either no weights  or  cluster size weights. This is most clearly seen when no HW SE correction has been used.

 We can also see that when SSDF  correction is used, then not using HW SE can result in small over-coverage especially for cluster-level non-adherence settings, which is improved when HW SE are used  (Figures \ref{fig:CluAdh_LargeCACE_Ybar_eligible} and \ref{fig:CluAdh_SmallCACE_Ybar_eligible}, 3$^{rd}$ and 5$^{th}$ rows of Panel A). When $J=50$ (Panel B), the use of SSDF-based distributions is not expected to make any material difference, and this is indeed the case. The impact of using HW SE  or the different weighting strategies is also minimal.

\subsection{Additional simulations}

Two extra additional scenarios are now considered to investigate the sensitivity of the CL-TSLS performance to number of clusters and cluster size imbalances, at both cluster and individual level adherence, but focusing on settings where confounding is strong with a large true LATE. 

In the first additional simulation, we explore the impact that the outcome ICC and the number of clusters have on bias, while leaving the expected total sample size fixed ($=1000$). 

We consider two marginal ICC for $Y_{ij}$ ($\rho_Y=0.05$ and $\rho_Y=0.80$) and three average cluster size ($n_j=20$, $10$ and $2.5$, corresponding to whether the number of clusters varied from $J=50, 100$ or $400$), which includes one of the scenarios previously considered in the main simulations for comparison. Though CRTs rarely have ICCs above $0.10$ \cite{murray2003methods},  the value of $\rho_Y=0.80$ is included to evaluate the performance of the methods in extreme settings.

In the second additional set of simulations we explore the effect of high cluster size imbalances.
While keeping the average sample size equal to $1000$, and $J=10$ or $50$, we create high cluster size imbalance using a Pareto distribution to generate  the cluster sizes \cite{guittet2006planning}.
The Pareto distribution parameters are chosen so that  approximately 40\% of the clusters have a size below 15, and 60\% a size above 15, while the average cluster size is $20$ and the minimum cluster size is $10$, resulting in approximately 1.8 for the shape and 9.1 for the scale.

\subsubsection{Results}

\begin{figure}
\thispagestyle{empty}
		\centering
		\includegraphics[scale=0.5]{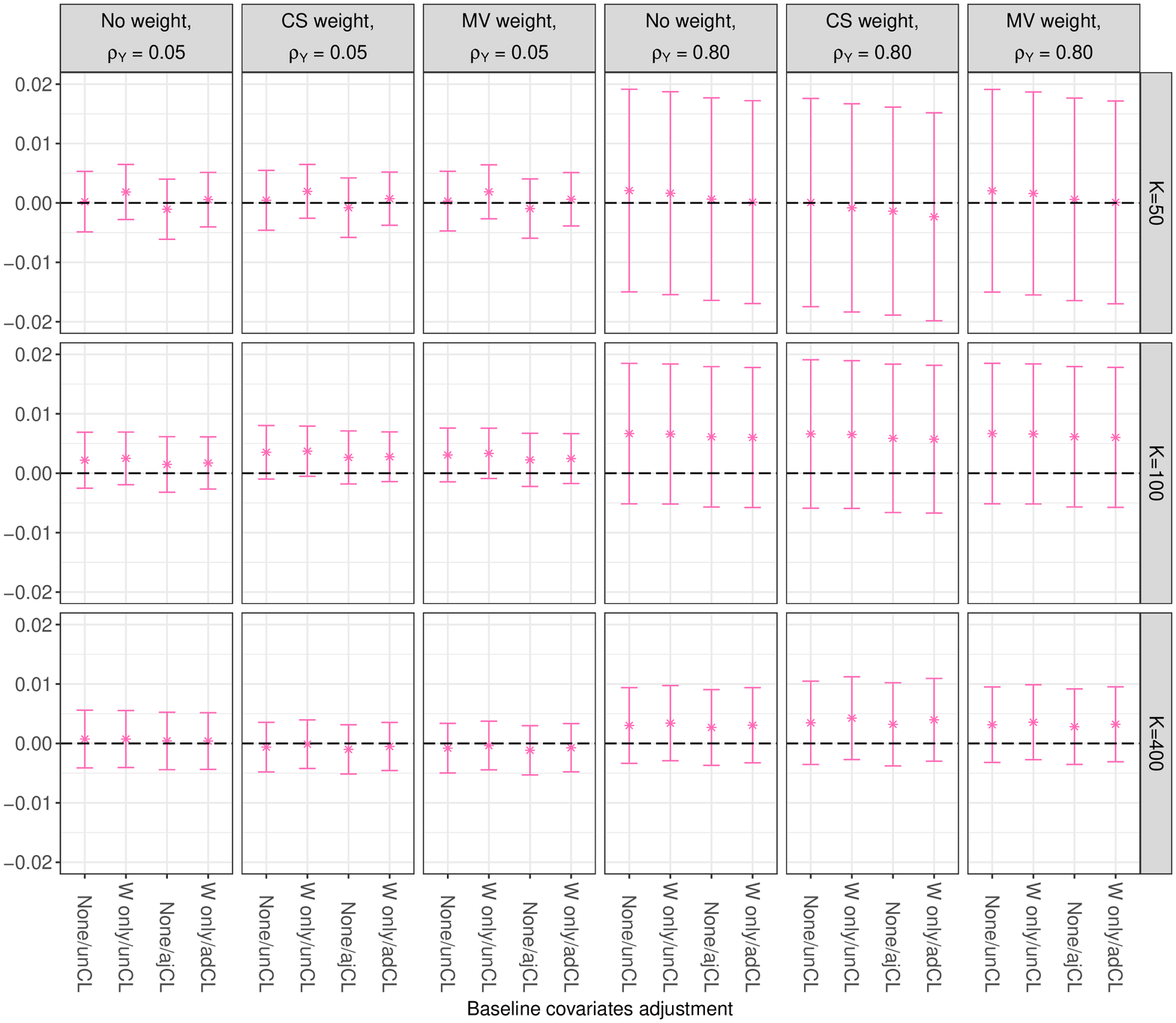}
		\caption{Bias of the CL-LATE for the extra simulation where non-adherence is at the cluster-level and a large true LATE, with high ICCs and varying numbers of clusters. Estimates are obtained via unadjusted or adjusted TSLS with different weights (none, cluster size (CS) and minimum-variance (MV)). Number of clusters varies by rows and ICC by column.
		}
	\label{fig:SmallSampleBias_CluAdh_LargeCACE_Ybar_eligible}
	\end{figure}

\begin{figure}
\thispagestyle{empty}
		\centering
		\includegraphics[scale=0.5]{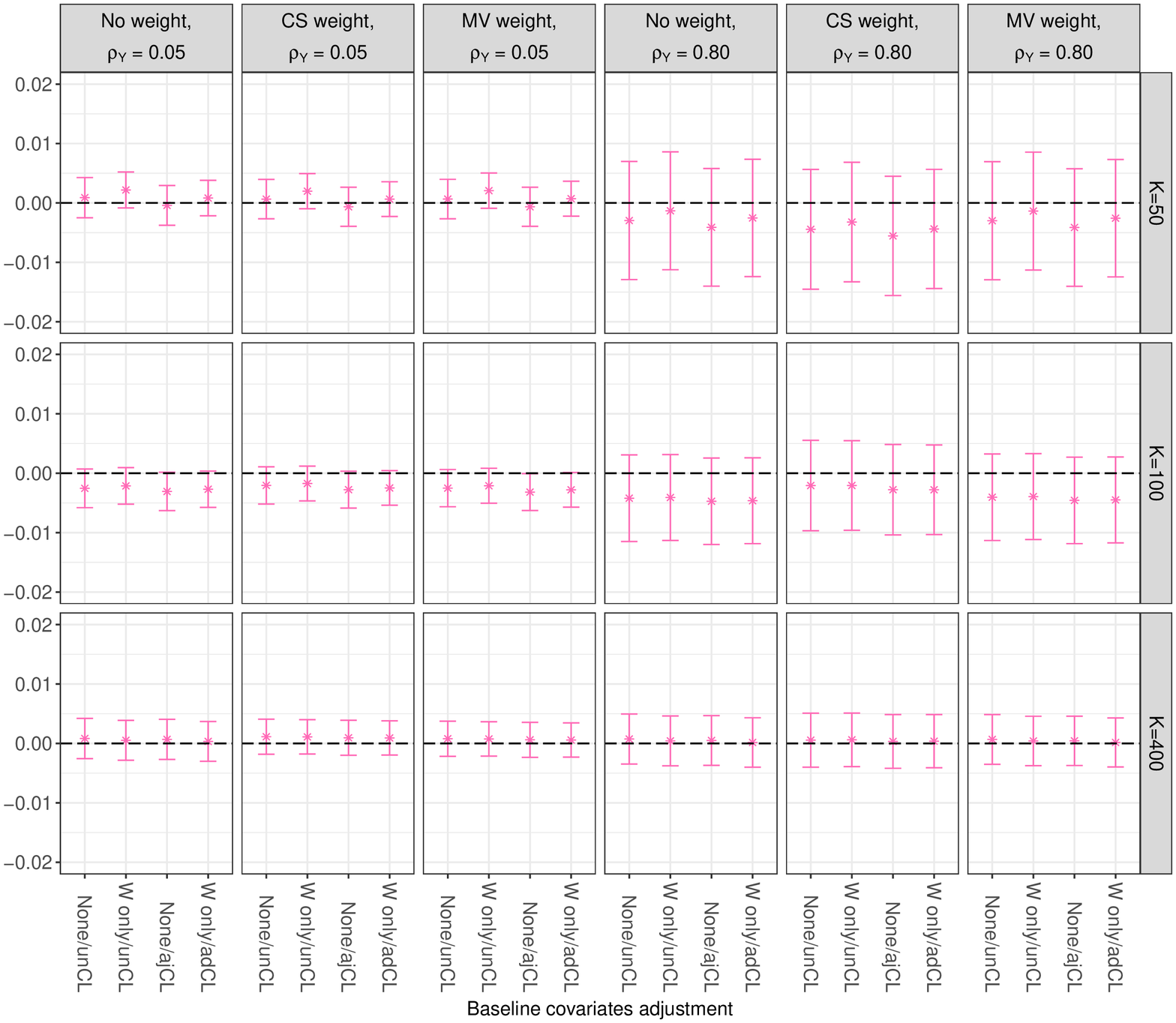}
		\caption{Bias of the CL-LATE for the extra simulation where non-adherence is at the individual-level and a large true LATE, with high ICCs and varying numbers of clusters. Estimates are obtained via unadjusted or adjusted TSLS with different weights (none, cluster size (CS) and minimum-variance (MV)). Number of clusters varies by rows and ICC by column.
		}
	\label{fig:SmallSampleBias_IndAdh_LargeCACE_Ybar_eligible}
	\end{figure}

Figures \ref{fig:SmallSampleBias_CluAdh_LargeCACE_Ybar_eligible} and \ref{fig:SmallSampleBias_IndAdh_LargeCACE_Ybar_eligible}, corresponding to cluster and individual level non-adherence settings,  show that for a fixed number of clusters (cells in the same row), the bias increases with increasing ICC for $Y$, but that as the number of clusters increase (moving down the column in the Figure), CL-TSLS results in negligible mean bias, even a very large $\rho_Y$. It is well known that TSLS is only asymptotically unbiased, and with CL analyses, we expect the asymptotics to depend on the number of clusters, and not the number of individuals. Nevertheless, the CL-summaries treated as outcomes for the two models involved in TSLS contain less ``information'' when the ICC is higher, which translates into a larger number of clusters being necessary for the bias to be negligible.

\begin{figure}
\thispagestyle{empty}
		\centering
		\includegraphics[scale=0.5]{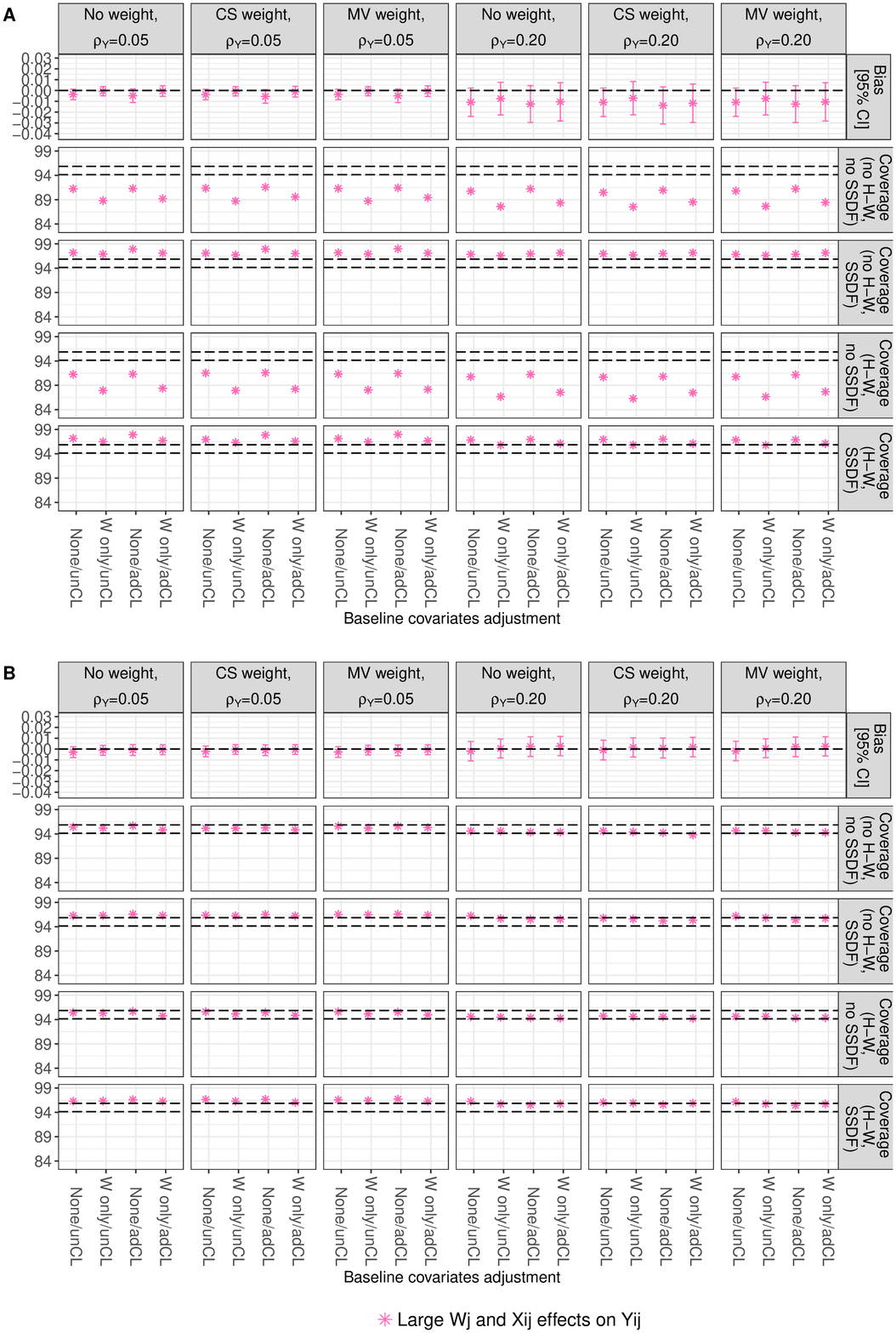}
		\caption{Extra simulation for very imbalanced cluster size settings. Bias (top row) and 95\% CI coverage  (Huber-White SEs (or not) and SSDF corrections (or not)) of the CL-LATE where non-adherence is at the cluster-level, and a large true LATE. Estimates are obtained via unadjusted or adjusted TSLS with different weights (none, cluster size (CS) and minimum-variance (MV)).  Small and large number of clusters results appears in Panel A and B respectively.
		}
		\label{fig:CluAdh_LargeCACE_Ybar_eligible_Pareto}
	\end{figure}

\begin{figure}
\thispagestyle{empty}
		\centering
		\includegraphics[scale=0.5]{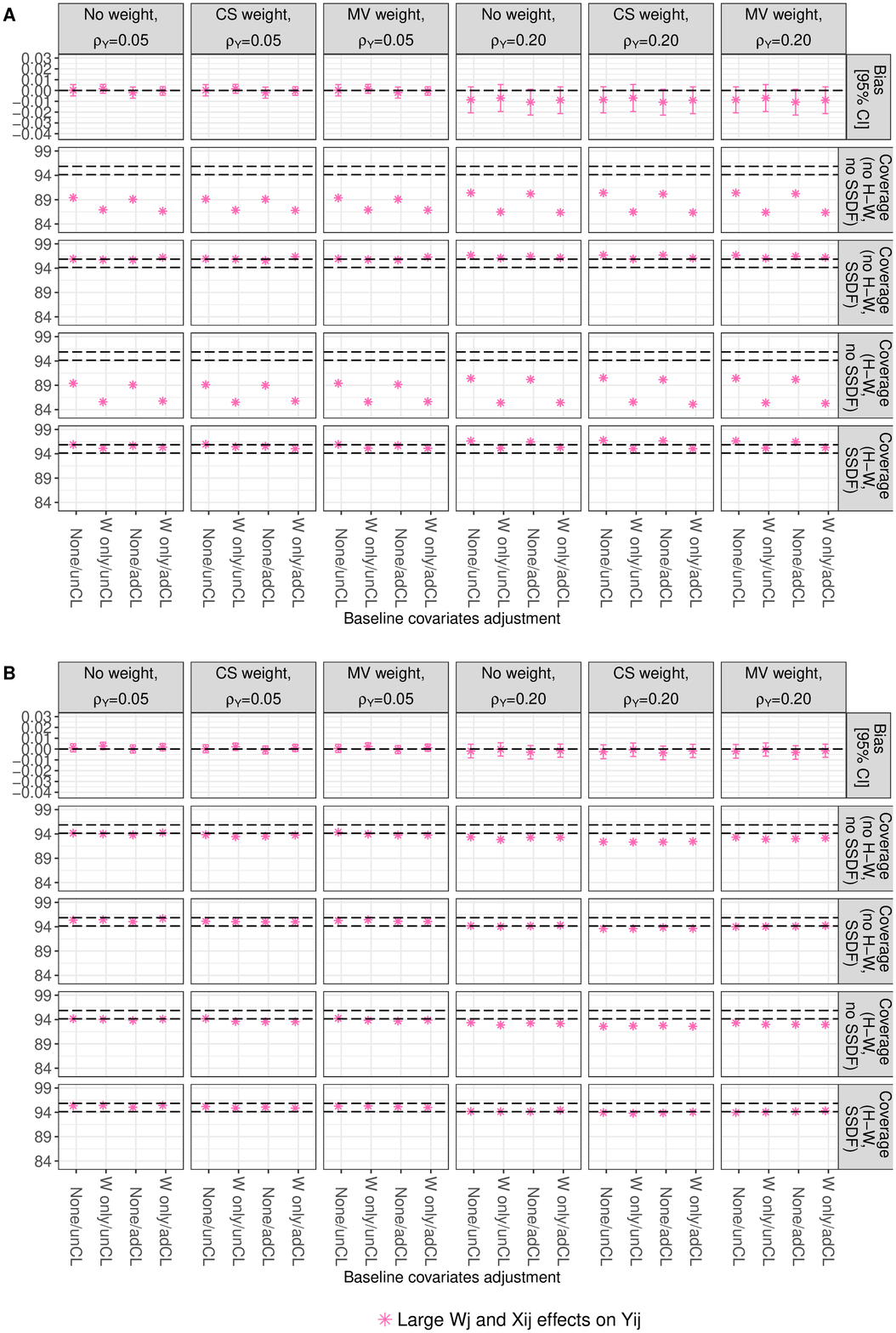}
		\caption{Extra simulation for very imbalanced cluster size settings. Bias (top row) and 95\% CI coverage  (Huber-White SEs (or not) and SSDF corrections (or not)) of the CL-LATE where non-adherence is at the individual-level, and a large true LATE. Estimates are obtained via unadjusted or adjusted TSLS with different weights (none, cluster size (CS) and minimum-variance (MV)).  Small and large number of clusters results appears in Panel A and B respectively.}
				
		\label{fig:IndAdh_LargeCACE_Ybar_eligible_Pareto}
	\end{figure}

The impact of high cluster size imbalance is reported in Figures \ref{fig:CluAdh_LargeCACE_Ybar_eligible_Pareto} and \ref{fig:IndAdh_LargeCACE_Ybar_eligible_Pareto}, where non-adherence is at the cluster and individual level respectively. We  see that when $J=10$ (Panel A), even with SSDF correction, failure to use HW SE results in under-coverage when using cluster size weights, which is especially pronounced when $\rho_Y$ is large. This is because cluster size weights are known to perform well when the cluster level residuals are homoscedastic, which is unlikely when cluster sizes are very imbalanced \cite{angrist2008mostly}.
This also explains why, using HW SE brings the coverage  close to nominal levels. This pattern is also observed  at $J=50$ (Panel B).

\section{Illustrative example} \label{Sec:TXT}

We now illustrate the methods in practice by applying each in turn to the analysis of the TXT4FLUJAB trial. This was a CRT of general practices in England aiming at estimating the effect of text messaging influenza vaccine reminders on increasing vaccine uptake in patients with chronic conditions, carried during the 2013 influenza season \cite{herrett2016text}. General practices (GPs) were stratified by the type of software used for text messaging and randomised to either standard care (control group, 79 GPs and $51 136$ patients) or a text messaging campaign (active group, 77 GPs and $ 51 121$ patients). Practices were not blinded to their allocation. GPs were the unit of analysis and the outcome of interest was the proportion of influenza vaccine uptake at the GP level. 

Influenza vaccination within the GPs  was automatically recorded in the clinical system from which the  data were extracted, so there are no missing data.

Since non-adherence was anticipated, the original statistical analysis plan specified obtaining by IV regression an efficacy estimate at the GP level \cite{herrett2016text}.
The original publication reported an estimated increase in vaccine uptake from texting reminders  of 14.3\% (95\% CI --0.59\% to 29.2\%) \cite{herrett2016text},
after dichotomising adherence at the cluster-level as either 100\% of eligible patients, compared with texting $<100\%$. 

Adherence to the intervention at the individual level could not be measured for all practices because it was recorded in a usable form only for GPs using a specific software.
Therefore, for these re-analyses, we restrict the dataset to 116 GPs  (58 in the intervention and 58 in the standard care arm) for which  individual-level adherence data are available.  Six of the 58 practices (10\%) in the intervention arm, did not send any reminders.
Conversely,  21 of the 58 practices (36\% in the standard care arm actually sent a  reminder to at least one patient.  Hence  non-adherence is two-sided. It also varies at the individual level. The median (range) of percentage of non-adherence at the GP level was 0\% (0\%-78.4\%) and 21.0\% (0\%-83.5\%) in the control and active group, respectively (Table \ref{BaselineTxt4Flu}).

The characteristics of the GPs and of the patients included in these analyses are comparable across trial groups  (Table \ref{BaselineTxt4Flu}); further the marginal ICC for individual-level outcome (vaccination) and treatment received (text message reminder) was 0.03 and 0.84 on the log-odds scale, respectively.

\begin{longtable}{lll}
	\caption{Baseline characteristics and percentages of non-adherence for the TXT4FLUJAB trial.} \label{BaselineTxt4Flu} \\
	\hline
	\noalign{\vskip 2mm}
	\multicolumn{1}{l}{\bf{Characteristics}} & \bf{Control} & \bf{Active} \\
	\endfirsthead
	\caption*{\textbf{Table 3 Continued}} \\
	\hline
	\multicolumn{1}{l}{\bf{Characteristics}} & \bf{Control} & \bf{Active } \\
	\midrule
	\endhead
	\bottomrule
	\multicolumn{1}{r@{}}{Continued on next page} \\
	\endfoot
	\endlastfoot
	\midrule
	
	\multicolumn{1}{l}{Practice-level characteristics} & & \\
	\multicolumn{1}{l}{\hspace{1em} Number of practices, n (\%)} &  58 (100.0) &  58 (100.0) \\
	\multicolumn{1}{l}{\hspace{1em} Open on weekends, n (\%)} & 39 (67.2) & 37 (63.8) \\
	\multicolumn{1}{l}{\hspace{1em} Patients per practice, median (range)} & 660 (148-$1\;678$) & 684 (79-$3\;022$) \\
	\noalign{\vskip 1mm}
	\multicolumn{1}{l}{Patient-level characteristics} & & \\
	\multicolumn{1}{l}{\hspace{1em} Number of patients, n (\%)}  & $40\;633$ (100) & $41\;073$ (100) \\
	\multicolumn{1}{l}{\hspace{1em} Male, n (\%)} & 20 752 (51.1) & 21 012 (51.2) \\
	\multicolumn{1}{l}{\hspace{1em} Has any disease, n (\%)} & $39\;244$ (96.6) & $39\;672$ (96.6) \\
	\multicolumn{1}{l}{\hspace{1em} Age, median (range)} & 50 (18-64) & 50 (18-64) \\
	\midrule
	
	\multicolumn{1}{l}{\bf{Active treatment received}} & & \\ 	
	\midrule 			
	\multicolumn{1}{l}{Patients receiving text message reminders, n (\%)} & $2\;628$ (6.5) & $11\;113$ (27.1) \\	
	\multicolumn{1}{l}{Practices sending text message reminders, n (\%)} & 21 (36.2) & 52 (80.7) \\		
	\multicolumn{1}{l}{\% of patients in each GP receiving reminders,} & 0 (0-78.4) & 21.0 (0-83.5) \\
	\multicolumn{1}{l}{ median (range)} & & \\

	\bottomrule
\end{longtable}

For our re-analysis, we  begin by discussing the plausibility of the identification assumptions. The  unconfoundedness assumption of the CL randomised treatment  is satisfied by design.
To check whether cluster randomisation is a relevant instrument,  we perform a test on the first stage of the CL-TSLS. The corresponding F-statistic is $F(1,114)=28.7>10$ thus passing  Staiger and Stock's rule of no null first-stage \cite{staiger1994instrumental}.

The exclusion restriction at the individual level implies that there is no other mechanism by which the GP being randomised to sending text vaccination reminders can affect a patient's actual vaccination uptake beside via the sending of the message. This assumption needs further justification, as in principle, a GP randomised to send reminders can be more conscious of the risks the patients face during the influenza season and use other means to remind at-risk patients, either in person, by post or by putting out flyers and posters in the clinic. So, it is possible that there  are patients who do not receive text reminders  and yet are prompted to get vaccinated by other means, by virtue of their practice being in the active group.
However, flyers, posters and postal letters already form part of regular care, so we believe they do not really vary by whether the GP is randomised to the active group.

The monotonicity assumption (that there  are no defiers)  also seems plausible as GPs randomised to the active group were more likely to send a text message reminder than those in the control group (see Table \ref{BaselineTxt4Flu}).

Finally,  there is a small risk of interference. The cluster  defined by GP practice should minimise this, as we only need to assume no interference at the cluster level, but it could be plausible that patients interact with those outside their GP, so that the exposure to a text message reminder of one patient may indeed affect the potential outcome, in this case, influenza vaccination of another patient from a different GP. The risk is small as usually close family members belong to the same general practice.

CL-TSLS on unadjusted CL outcomes was implemented adjusting and not adjusting for a baseline CL covariate, namely whether the  clinic was open on the weekends (yes/no).
Table \ref{Txt4Flu}  shows the CL-LATE estimates (expressed as mean risk differences), with  95\% CIs  and p-values obtained via different weighting strategies, and corrections.

 Using cluster size weights results in different point estimates from the rest. This was expected as there is substantial cluster size imbalance (cluster size range: $148$--$ 1 678$ in the control group and $79$--$3022$ in the active group  (Table \ref{BaselineTxt4Flu}).  The results obtained using no weights or MV weights leads to point estimates that are very close to those found  in the original  publication \cite{herrett2016text}.

In terms of inference, the use of SSDF correction in calculating CIs is not important, as the number of clusters is large, but the HW SEs paired with MV weighting provides efficiency gains, especially for the adjusted  CL-TSLS analyses. Overall however, the CIs are still very wide.

These results suggest that there is weak evidence that  receiving a text reminder increases the expected proportion of patients within a compliant practice that get vaccinated against influenza  by 14\%  (95\% CI: $-0.5$ to  $30$\%, $p=0.065$, based on the adjusted CL-TSLS using MV weights and normal-based CI with HW SEs estimate).

Contrast this with the  unadjusted CL-summaries mean risk differ zence ITT estimate,  which  indicates a 2.89\% increase (95\% CI $-0.17$ to $5.95$, $p=0.064)$, highlighting the dilution effects deriving from the non-adherence.

One of the disadvantages of TSLS is lack of efficiency. Adjusting for individual-level baseline covariates may  help obtaining narrower CIs. Since CL-TSLS cannot adjust for individual level covariates, we now perform the analyses using  adCL summary outcomes, generated by adjusting for gender, age and the presence of disease. Results are reported in Table \ref{Txt4Flu_adCL} in the Appendix.   The results do not materially change (weak evidence of a 13\% increase vaccination uptake), possibly  because these individual level covariates are not strongly associated with the outcome.

Our illustrative example is limited by  the availability of baseline cluster-level  variables. Since there was only one CL-variable recorded, the impact of covariate adjustment on the CL-TSLS is negligible.  Other limitations of these results include the possibility of measurement error, for if patients received their influenza vaccine outside the practice, this would not have been recorded in the system, unless the patient informed their GP.

\begin{small}
	\begin{center}
		\begin{longtable}{llllll}
			\caption{LATE of text message reminders to receive flu vaccination on the uptake of flu vaccine in the TXT4FLUJAB trial, using unadjusted CL-summaries} \label{Txt4Flu} \\
			\hline
			\hline
			\noalign{\vskip 0.5mm}
\multicolumn{1}{l}{{\textbf{}}} & \multicolumn{1}{l}{{\textbf{}}} & \multicolumn{2}{c}{{\textbf{Unadjusted}}} & \multicolumn{2}{c}{{\textbf{Adjusted$^\text{a}$}}} \\
			\multicolumn{1}{l}{{\textbf{}}} & \multicolumn{1}{l}{{\textbf{}}} & \multicolumn{1}{l}{{\textbf{LATE (95\% CI)}}} & \multicolumn{1}{l}{{\textbf{p}}} & \multicolumn{1}{r}{{\textbf{LATE (95\% CI)}}} & \multicolumn{1}{c}{{\textbf{p}}} \\
			\noalign{\vskip 0.5mm}
			\endfirsthead
			\caption*{\textit{Table \ref{Txt4Flu} Continued}} \\
			\hline
			\noalign{\vskip 0.5mm}
\multicolumn{1}{l}{{\textbf{Weighting}}} & \multicolumn{1}{l}{{\textbf{SE \&}}} & \multicolumn{2}{c}{{\textbf{Unadjusted}}} & \multicolumn{2}{c}{{\textbf{Adjusted$^\text{a}$}}} \\
			\multicolumn{1}{l}{{\textbf{strategy}}} & \multicolumn{1}{l}{{\textbf{correction}}} & \multicolumn{1}{l}{{\textbf{LATE (95\% CI)}}} & \multicolumn{1}{l}{{\textbf{p}}} & \multicolumn{1}{r}{{\textbf{LATE (95\% CI)}}} & \multicolumn{1}{c}{{\textbf{p}}} \\
			\noalign{\vskip 0.5mm}
			\midrule
			\endhead
			\bottomrule
			\multicolumn{4}{r@{}}{\it Continued on next page} \\
			\endfoot
			\endlastfoot
			\midrule
			No weighting   		& None               & 0.149 (-0.006,0.305) & 0.060 & 0.148 (-0.078,0.303) & 0.063 \\
			 				& HW        & {\color{white} 0.149} (-0.006,0.305) & 0.060 & {\color{white} 0.148} (-0.005,0.301) & 0.058 \\
							& SSDF 				 & {\color{white} 0.149} (-0.009,0.308) & 0.065 & {\color{white} 0.148} (-0.012,0.308) & 0.069 \\
							& SSDF + HW & {\color{white} 0.149} (-0.009,0.308) & 0.065 & {\color{white} 0.148} (-0.009,0.305) & 0.064 \\
			\noalign{\vskip 1.0mm}
			Cluster size 		& None	             & 0.071 (-0.065,0.207) & 0.307 & 0.074 (-0.061,0.209) & 0.284 \\
			 weights			& HW        & {\color{white} 0.071} (-0.088,0.230) & 0.382 & {\color{white} 0.074} (-0.077,0.225) & 0.338 \\
							& SSDF				 & {\color{white} 0.071} (-0.068,0.209) & 0.313 & {\color{white} 0.074} (-0.064,0.212) & 0.292 \\
							& SSDF + HW & {\color{white} 0.071} (-0.091,0.233) & 0.388 & {\color{white} 0.074}  (-0.081,0.228) & 0.346 \\
			\noalign{\vskip 1.0mm}
			Minimum-   		& None               & 0.143 (-0.008,0.293) & 0.064 & 0.142 (-0.009,0.293) & 0.065 \\
			variance	& HW        & {\color{white} 0.143} (-0.006,0.291) & 0.060 & {\color{white} 0.142} (-0.005,0.289) & 0.058 \\
				 weights			& SSDF 				 & {\color{white} 0.143} (-0.011,0.296) & 0.069 & {\color{white} 0.142} (-0.012,0.297) & 0.071 \\
							& SSDF + HW & {\color{white} 0.143} (-0.009,0.294) & 0.065 & {\color{white} 0.142} (-0.008,0.293) & 0.064 \\
			\bottomrule
			\bottomrule
			\multicolumn{6}{l}{$^\text{a}$ \footnotesize{Adjusted for whether clinic is opened during} weekends.} \\
			\multicolumn{6}{l}{\footnotesize{HW: Huber-White; SSDF: small sample degrees of freedom.}}
		\end{longtable}
	\end{center}
\end{small}

\section{Discussion}\label{Sec:Discussion}

This paper demonstrates the use of TSLS regression applied to CL summaries as a simple and valid method for obtaining estimates of the LATE in CRTs where non-adherence occurs at either the cluster or the individual level. The performance of CL-TSLS with different weighting strategies (none, cluster size, minimum variance) and methods for obtaining CIs (alternatively using or not Huber-White SEs and/or SSDF correction).

We have demonstrated empirically through simulations that under the stated sufficient assumptions for identification TSLS regression of CL summaries provides consistent estimates of the causal treatment effect in the sub-population of compliers, where non-adherence  is at the cluster level.  With individual-level non-adherence, the additional assumption that the cluster-specific LATE is homogeneous across clusters  is required for CL-TSLS to identify the population LATE \cite{Kang2018}. Moreover, provided that an appropriate distribution with SSDF adjustment is used when the number of clusters is small and Huber-White SEs are used if there is high cluster size imbalance,  valid 95\% CIs can be constructed.

Our simulation study suggests that all weighting  strategies perform similarly when the number of clusters is not small. When the number of clusters is small, MV weights tend to be badly estimated, and are not recommended; furthermore when the cluster sizes are very variable, cluster size weights should not be used.  Although in the simulations the choice of weights did not affect the point  estimates, these were affected in the illustrative example. Overall our results show that, unless there are very few clusters, or the outcome ICC is large,  MV weighting  performs well \cite{kerry2001unequal}.

 Although CL-TSLS is easy to implement, it suffers from being very inefficient. 
 We can see this in the illustrative example  where all CIs for the CL-TSLS LATE estimates are much wider than those for the estimated ITT. There are two reasons for this: CL-analyses are inefficient, unless the cluster sizes are (almost) equal \cite{donner1998some},  and TSLS is known to be inefficient, although adjusting  for baseline covariates can ameliorate this \cite{Wooldridge2010}. In the context of CL-analysis it is only possible to include cluster-level baseline covariates in the regressions \cite{angrist2008mostly}. However, we tested the performance of  CL outcome  which are adjusted for individual-level covariates \cite{hayes2009cluster}, and showed that this indeed has the potential to improve efficiency in certain settings.

For CL-TSLS analyses, inference should be based on the number of clusters, with CIs constructed by using  $t-$ distributions with DFs equal to $J-p$ \cite{donald2007inference}. The outcome ICC value is important too, with higher ICCs requiring a larger number of clusters for the asymptotical arguments to work, as well as whether the cluster-level variances are homoscedastic \cite{angrist2008mostly}.  

Other methods for estimating causal treatment effects in CRTs with non-adherence at the individual level exist, in particular Kang and Keele \cite{Kang2018} have recently proposed a finite-sample estimator that identifies the population LATE and obtains  valid inferences even when compliance is low.

We do not consider here situations where the identification assumptions are violated.
There are several options to study the sensitivity to departures from these  assumptions.  For example, if the exclusion restriction does not hold, a Bayesian parametric model can use priors on the non-zero direct effect of  randomisation on the outcome for identification \cite{Conley2012}. Since the models are only weakly identified,  the results depend strongly on the  prior distributions.
Alternatively, violations of the exclusion restriction can also be handled by using baseline covariates to model the probability of compliance directly, within structural equation modelling via expectation-maximisation framework \cite{Jo2002, Jo2002a}.

We have only focused on LATE estimands. These are often criticised because the estimates obtained apply to the ``compliers'' in the population, and these cannot be observed in practice, thus limiting applicability. However, LATE estimates may be used to provide  information about the average causal effect in the entire population   \cite{baiocchi2014instrumental}.
Moreover, the average treatment effect on the compliers is often of interest to patients and medical decision makers,  especially  when they expect patients to comply with the treatment \cite{Murray2018}.

\begin{small}
\textbf{Acknowledgements}\\
We thank the TXT4FLUJAB study team for access to the data. 
SCA was funded by a UK Economic and Social Research Council PhD scholarship ES/J500021/1. KDO was supported by UK Medical Research Council Career development award in Biostatistics MR/L011964/1.

\textbf{Conflict of interest}\\
The Authors declare that there is no conflict of interest.
\end{small}

\clearpage

	\clearpage

\clearpage

\clearpage
\clearpage

\clearpage

\begin{appendices}
\section{Adjusted cluster-level summaries for binary data} \label{appendixA}	

For binary a standard logistic regression model is usually fitted for binary outcomes, which assumes that
\begin{equation}
\text{logit}\left( \pi_{ij}\right)=\log\left( \frac{\pi_{ij}}{1-\pi_{ij}}\right) = \lambda_1+\lambda_2X_{ij}
\label{model_firststage}
\end{equation}
Let $M_{j} $ and $ \hat{M}_{j} $ be the observed and predicted number of successes in the $j$th cluster, respectively. After fitting model (\ref{model_firststage}), $ \hat{M}_{j} $ is calculated as
\begin{equation}\nonumber
\hat{M}_{j} = \sum_{l=1}^{m}\hat{\pi}_{ij} = \sum_{i=1}^{n_j} \text{expit} \left( \hat{\lambda}_1+\hat{\lambda}_2X_{ij} \right).
\end{equation}
Then the observed and predicted numbers of success are compared by computing a residual for each cluster.  If we want to estimate the adjusted RD, the residual, known as difference-residual,  for each cluster is calculated as $$e_{j}={(M_{j}-\hat{M}_{j})}/{n_j},$$ and treated as a continuous outcome in any subsequent analyses.

\section{Performance criteria}
Let  the  mean of the estimated LATE across the replicate datasets in each scenario, indexed by $l=1,...,L$, with $L=2,500$ be $\bar{\hat{\beta}}_{IV}=\frac{1}{L}\sum\limits_{l=1}^{L} \hat{\beta}_{IV_l}$. 
The following criteria  were used to assess the performance of the methods investigated 

\begin{enumerate}[(a)]
\item \textbf{Empirical bias:}   estimated by $\bar{\hat{\beta}}_{IV}- \beta_{CZ}$. 
\item \textbf{Monte Carlo error (MCE) of empirical bias} $= \sqrt{\sum_{l=1}^{L} \left(\hat{\beta}_{IV_l}-\bar{\hat{\beta}}_{IV}\right)^2/[L(L-1)]}$.
\item \textbf{Coverage rate of the nominal of 95\% CIs} $\frac{1}{L}\sum_{l=1}^{L} I \Big(|\hat{\beta}_{IV_l} - \beta_{CZ}| < 1.96 s_i\Big)$, where $s_i$ denotes the model-based standard error for $\hat{\beta}_{IV_l}$.
The MCE of coverage is $\sqrt{\sum_{l=1}^{L} (0.95)(0.05)/L}$,which means that the expected range of the nominal 95\% CIs is between 94.1\% and 95.9\%. 
\end{enumerate}

\pagebreak

\section{Results for adjusted CL summaries CL-TSLS for TEXT4FLUJAB}
\begin{small}
	\begin{center}
		\begin{longtable}{llrlrl}
			\caption{TSLS estimation of practice-level LATE of reminder text messaging to receive flu vaccine on the percentage uptake of flu vaccine in the TXT4FLUJAB trial using  adjusted CL outcomes, adjusting for individual-level covariates gender, age and presence of disease.} \label{Txt4Flu_adCL} \\
			\hline
			\hline
			\noalign{\vskip 0.5mm}
\multicolumn{1}{l}{{\textbf{}}} & \multicolumn{1}{l}{{\textbf{}}} & \multicolumn{2}{c}{{\textbf{Unadjusted}}} & \multicolumn{2}{c}{{\textbf{Adjusted$^\text{a}$}}} \\
			\multicolumn{1}{l}{{\textbf{}}} & \multicolumn{1}{l}{{\textbf{}}} & \multicolumn{1}{l}{{\textbf{LATE (95\% CI)}}} & \multicolumn{1}{l}{{\textbf{p}}} & \multicolumn{1}{r}{{\textbf{LATE (95\% CI)}}} & \multicolumn{1}{r}{{\textbf{p}}} \\
			\midrule
			No weighting   		& None             & 0.133 (-0.016,0.282) & 0.081 & 0.133 (-0.017,0.282) & 0.082 \\
			 				& HW       &  (-0.016,0.282) & 0.081 &  (-0.014,0.280) & 0.077 \\
							& SSDF 				 &  (-0.019,0.285) & 0.086 &  (-0.021,0.286) & 0.089 \\
							& SSDF + HW & (-0.019,0.285) & 0.086 &  (-0.018,0.283) & 0.083 \\
			\noalign{\vskip 1.0mm}
			Cluster 		& None	             & 0.068 (-0.063,0.198) & 0.310 & 0.071 (-0.058,0.200) & 0.280 \\
			size	 weighting 		& Huber-White        & (-0.081,0.216) & 0.372 & (-0.069,0.212) & 0.320 \\
							& SSDF				 &  (-0.065,0.201) & 0.316 & (-0.061,0.203) & 0.288 \\
							& SSDF + HW &  (-0.084,0.219) & 0.378 &  (-0.073,0.215) & 0.328 \\
			\noalign{\vskip 1.0mm}
			Minimum-   		& None               & 0.128 (-0.017,0.273) & 0.084 & 0.128 (-0.017,0.273) & 0.084 \\
			variance weighting 		& HW      & (-0.015,0.271) & 0.080 &  (-0.014,0.269) & 0.077 \\
							& SSDF 				 & (-0.020,0.275) & 0.090 &  (-0.021,0.277) & 0.091 \\
							& SSDF + HW & (-0.018,0.273) & 0.086 & (-0.017,0.273) & 0.083 \\
			\bottomrule
			\bottomrule
			\multicolumn{6}{l}{$^\text{a}$ \footnotesize{TSLS estimation was adjusted for weekend clinics (yes/no).}}
		\end{longtable}
	\end{center}
\end{small}

\section{Results for small true LATE}
	\clearpage
\begin{figure}
\thispagestyle{empty}
	\centering
		\includegraphics[scale=0.5]{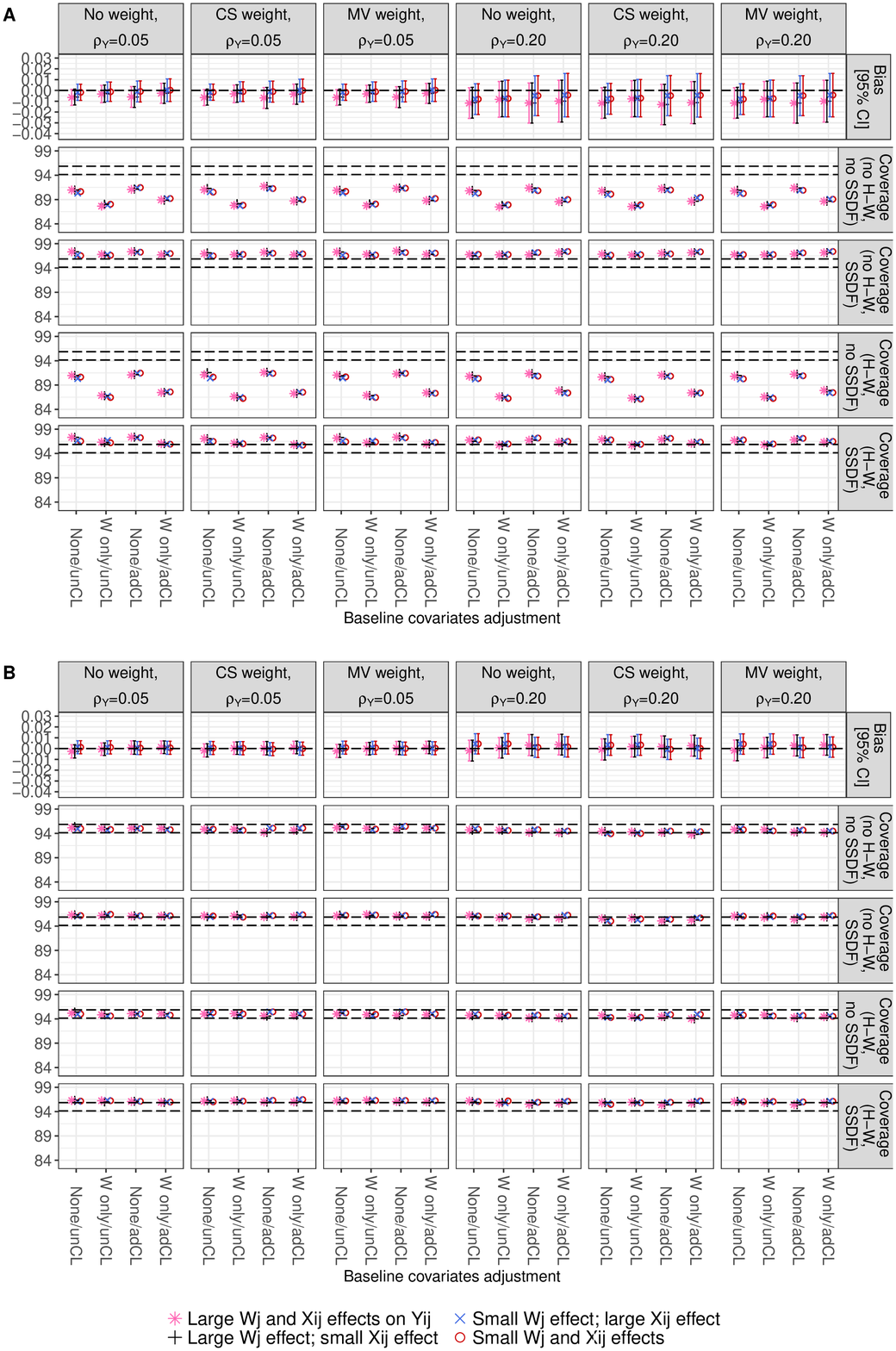}
\caption{\small{Bias (top row) and 95\% CI coverage (rows 2--5) of CL-LATE with cluster-level non-adherence and  small true LATE. Data generation scenarios represented by $\ast, +, \times$, and $\circ$. Estimates are obtained via unadjusted or $W$-adjusted TSLS with different weights (none, cluster size (CS) and minimum-variance (MV)) (by column) using  CL unadjusted or adjusted for $X$ outcomes (``unCL'' or ``adCL''). Small  ($J=10$) and large ($J=50$) number of clusters results are shown in Panel A  and B .}}		
\label{fig:CluAdh_SmallCACE_Ybar_eligible}
	\end{figure}

	\clearpage

\end{appendices}

\end{document}